\documentclass[aps,twocolumn]{revtex4}
\pdfoutput=1

\usepackage{graphicx}
\usepackage{times}
\usepackage{amsmath}
\usepackage{amssymb}

\begin{document}
\title{Transition states in protein folding}

\author{\hspace*{-1cm} Thomas R.\ Weikl\\[0.2cm]
\hspace*{-1cm} \small Max Planck Institute of Colloids and Interfaces, Department of Theory 
\\[-0.1cm] 
\hspace*{-1cm} \small and Bio-Systems, Science Park Golm, 14424 Potsdam, Germany}

\begin{abstract}
The folding dynamics of small single-domain proteins is a current focus of simulations and experiments. Many of these proteins are `two-state folders', i.e.~proteins that fold rather directly from the denatured state to the native state, without populating metastable intermediate states. A central question is how to characterize the instable, partially folded conformations of two-state proteins, in particular the rate-limiting transition-state conformations between the denatured and the native state. These partially folded conformations are short-lived and cannot be observed directly in experiments. However, experimental data from detailed mutational analyses of the folding dynamics provide indirect access to transition states. The interpretation of these data, in particular the reconstruction of transition-state conformations, requires simulation and modeling. The traditional interpretation of the mutational data aims to reconstruct the degree of structure formation of individual residues in the transition state, while a novel interpretation aims at degrees of structure formation of cooperative substructures such as $\alpha$-helices and $\beta$-hairpins. By splitting up mutation-induced free energy changes into secondary and tertiary structural components, the novel interpretation resolves some of the inconsistencies of the traditional interpretation. 
\end{abstract}

\maketitle

%%%
\section{Folding dynamics of small single-domain proteins } 
%%%
\label{section_folding_dynamics}

\begin{figure}[b]
\begin{center}
\resizebox{0.6\linewidth}{!}{\includegraphics{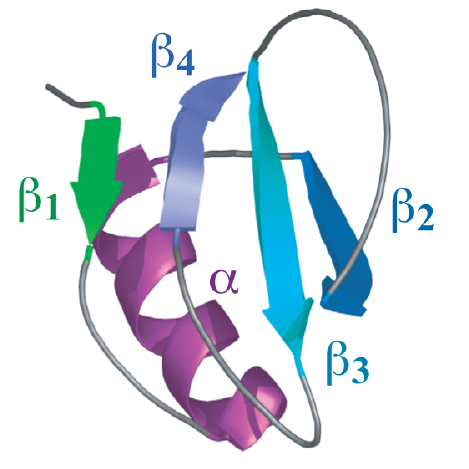}}
\end{center}
\caption{The structure of the protein CI2 consists of an $\alpha$-helix packed against a four-stranded $\beta$-sheet \cite{McPhalen87}. CI2 is a two-state protein that folds from the denatured state to the native state without experimentally detectable intermediate states \cite{Jackson91}.}
\label{figure_CI2structure}
\end{figure}

Proteins are biomolecules that participate in all cellular processes of living organisms. Some proteins have structural or mechanical function, such as the protein collagen, which provides the structural support of our connective tissues. Other proteins catalyze biochemical reactions, transport or store electrons, ions, and small molecules, perform mechanical work in our muscles, transmit information within or between cells, act as antibodies in immune responses, or control the expression of genes and, thus, the generation of other proteins \cite{Alberts02}. Proteins achieve this functional versatility by folding into different, unique three-dimensional structures (see fig.~\ref{figure_CI2structure}). The folding of proteins is a spontaneous process of structure formation and a prerequisite for their robust function. Misfolding can lead to protein aggregates that cause severe diseases, such as Alzheimer's, Parkinson's, or the variant Creutzfeldt-Jakob disease \cite{Dobson03}. 

How precisely proteins fold into their native, three-dimensional structure remains an intriguing question \cite{Dill07,Dill08}. Given the vast number of unfolded conformations of the flexible protein chain, Cyrus Levinthal argued in 1968 \cite{Levinthal68,Levinthal69} that proteins are guided to their native structure by a sequence of folding intermediates. In the following decades, experimentalists focused on detecting and characterizing metastable folding intermediates of proteins \cite{Baldwin99a}. The view that proteins have to fold in sequential pathways from intermediate to intermediate, now known as `old view' \cite{Baldwin94,Matthews93}, changed in the '90s when statistical-mechanical models demonstrated that fast and efficient folding can also be achieved on funnel energy landscapes that are smoothly biased towards the native state \cite{Dill97,Bryngelson95}. The stochastic folding process on these landscapes is highly parallel, and partially folded states along the parallel folding routes are instable rather than metastable. The paradigmatic proteins of this `new view' are two-state proteins, first discovered in 1991 \cite{Jackson91}. Two-state proteins fold from the denatured state to the native state without experimentally detectable intermediate states. Since then, the majority of small single-domain proteins with a length up to 100 or 120 amino acids has been shown to fold in apparent two-state kinetics, while larger multi-domain proteins often exhibit metastable folding intermediates  \cite{Jackson98,Fersht99,Grantcharova01}.

%Figure 

The simplest model for a two-state process is classical transition-state theory. In transition-state theory, the folding rate of a two-state protein is assumed to have the form (see, e.g., \cite{Fersht99})
\begin{equation}
k = k_o \exp[-G_\text{T-D}/RT] \label{kf_trans_state_theo}
\end{equation}
where $G_\text{T-D}$ is the free-energy difference between the transition state T and the denatured state D (see fig.~\ref{trans_state_theo}(a)), and $k_o$ is a prefactor that depends on the conformational diffusion coefficient of the protein. Classical transition-state theory thus assumes a third state, the transition state T, that governs the folding kinetics. From a statistical-mechanical perspective of protein folding, the transition state T, the denatured state D, and the native state N are ensembles of conformations. The denatured state is a huge ensemble of largely unstructured protein 
conformations, while the folded, native state corresponds to a rather narrow ensemble that captures the thermal fluctuations in this state. 
The transition state can be defined as an ensemble of partially folded conformations with equal probability to fold or unfold  \cite{Du98,Hummer04,Snow06}. According to this definition, a trajectory that passes through a transition-state conformation thus has the same probability 0.5 to proceed to the native state or to the denatured state from this conformation.

The folding times of small single-domain proteins range from microseconds to seconds \cite{Jackson98,Grantcharova01,Maxwell05}. An important observation was that these folding times correlate with the average `localness' of contacts between amino acids in the folded state \cite{Plaxco98,Plaxco00}. A local contact is a contact between two amino acids that are close in sequence, for example a contact between two amino acids in adjacent turns of an $\alpha$-helix.  Proteins with predominantly local contacts, such as $\alpha$-helical proteins, tend to fold faster than proteins with many nonlocal, sequence-distant contacts. The physical principle that underlies this correlation between folding times and average localness of contacts seems to be loop closure \cite{Weikl08a,Weikl03a}, since local contacts can be formed by fast closure of small loops \cite{Fersht00,Zhou04}.

Molecular dynamics (MD) simulations with detailed, atomistic models of proteins have been used to study the dynamics of small, fast-folding proteins with folding times in the microsecond range \cite{Ensign07,Duan98,Ferrara00,Snow02,Lei07,Freddolino08}. One of the best-studied proteins is the villin headpiece, an $\alpha$-helical protein with 36 amino acids. Central questions are whether folding simulations with current force fields reach the correct, experimentally determined folded state of a protein from unfolded conformations, and whether the dynamics of folding events observed in these simulations agrees with experimental data. In case of the villin headpiece, MD simulations of several groups have reached the folded state of the protein \cite{Ensign07,Lei07}, whereas folding simulations of a fast-folding WW domain, a $\beta$-sheet protein, have only reached structures with incorrect topology \cite{Freddolino08}. 

%%%
\section{Mutational analysis of two-state protein folding}
%%%
\label{section_mutational_analysis}

Since transition-state conformations of two-state proteins are instable and, thus, short-lived, they cannot be observed directly in experiments. The most important, indirect experimental method to investigate the folding dynamics of two-state proteins is mutational analysis \cite{Fersht99}. In a mutational analysis, a large number of mostly single-residue mutants of a protein is generated, and the folding rate $k$ and stability $G_\text{N-D}$ of each mutant is determined. The stability $G_\text{N-D}$ of a protein is the free energy difference between native state N and the denatured state D.  

%Figure 
\begin{figure}[t]
\begin{center}
\resizebox{\linewidth}{!}{\includegraphics{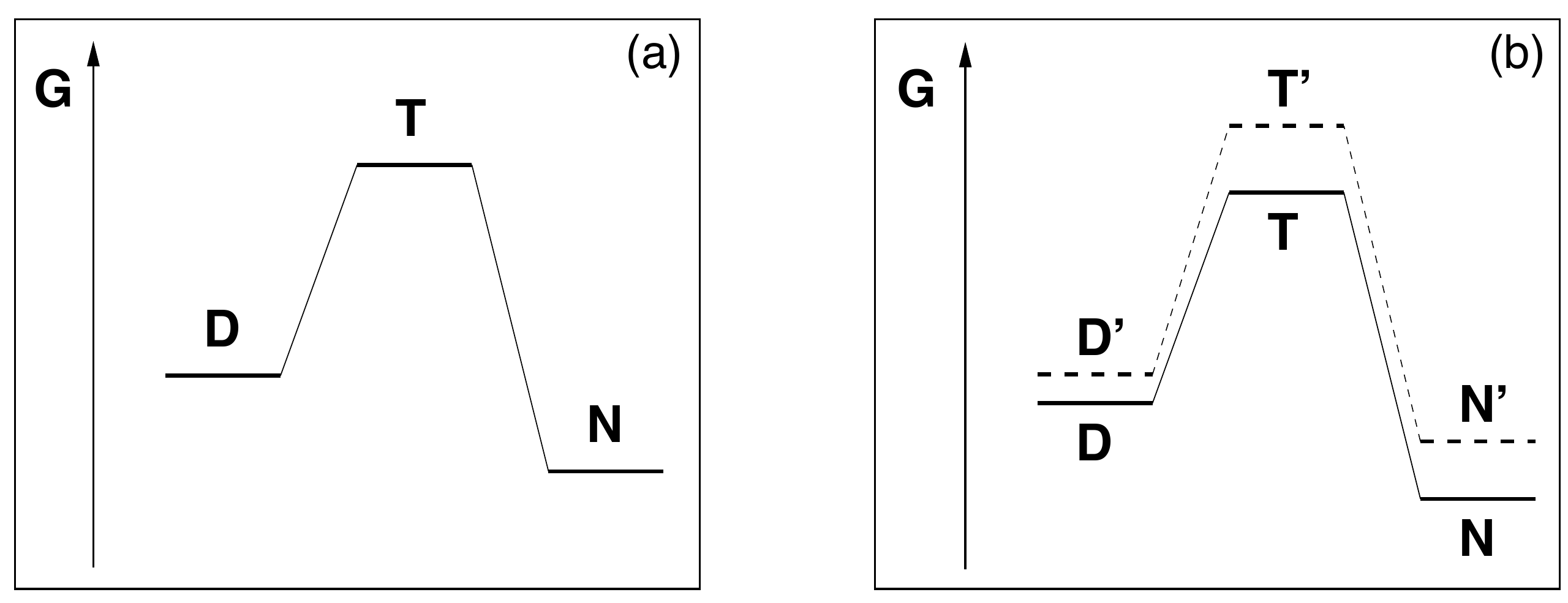}}
\end{center}
\caption{(a) In classical transition-state theory, the folding kinetics of a two-state protein is dominated by a transition state T between the denatured state D and the native state N.  The folding rate depends on the difference $G_\text{T-D}=G_\text{T}-G_\text{D}$ between the free energy $G_\text{T}$ of the transition state T and the free energy $G_\text{D}$ of the  denatured state D, see eq.~(\ref{kf_trans_state_theo}).  -- (b) Mutations perturb the free energies of the denatured state, transition state, and native state.}
\label{trans_state_theo}
\end{figure}

The effect of each mutation on the folding dynamics is typically quantified by its   
$\Phi$-value \cite{Matouschek89,Fersht99}
\begin{equation}
\Phi= \frac{R T\ln( k/k')}{\Delta G_\text{N-D}} \label{in_phi_def}
\end{equation}
Here, $k$ is the folding rate for the wildtype protein, $k'$ is the folding rate for the mutant protein, and $\Delta G_\text{N-D}=G_\text{N'-D'}-G_\text{N-D}$ is the change of the protein stability induced by the mutation.  $G_\text{N'-D'}$ and $G_\text{N-D}$ denote the stabilities of the mutant and the wildtype, see fig.~\ref{trans_state_theo}(b).  With eq.~(\ref{kf_trans_state_theo}), $\Phi$-values can be written in the form 
\begin{equation}
\Phi=\frac{\Delta G_\text{T-D}}{\Delta G_\text{N-D}}  \label{in_phi_trans_state_theo}
\end{equation} 
if one assumes that the pre-exponential factor $k_o$ is not affected by the mutation \cite{Fersht99}. Here, $\Delta G_\text{T-D}= G_\text{T'-D'}- G_\text{T-D}$ is the mutation-induced change of the free-energy barrier $G_\text{T-D}$, see fig.~\ref{trans_state_theo}(b). 

In the past decade, the folding dynamics of several dozen two-state proteins has been investigated with mutational $\Phi$-value analyses (for references, see, eg.~\cite{Weikl07}). An example of data from a mutational analysis of the protein CI2 \cite{Itzhaki95} is shown in table 1. The single-residue mutations of table 1 are all located in the $\alpha$-helix of the protein CI2, which comprises the residues 12 to 24 of this protein (see fig.~\ref{figure_CI2structure}). In the mutation S12G, for example, the amino acid 12 of the wildtype, Serine (single-letter code S) is replaced by the smaller amino acid Glycine (single-letter code G). The experimentally measured $\Phi$-value for this mutation is 0.29, and the experimentally measured change in stability $\Delta G_N$ is 0.8 kcal/mol.

% Table 
\begin{table}
\caption{Mutational data for the helix of the protein CI2 ~~~~~~~~~~~~~~~~~~~~~~~}
\label{ts_data_CI2helix}
\begin{center}
\begin{tabular*}{\columnwidth}{@{\extracolsep{\fill}}cccc}
mutation & $\Phi$ & $\Delta G_N$   & $\Delta G_\alpha$  \\ 
\hline
S12G & 0.29 & 0.8 & 0.28 \\
S12A & 0.43 & 0.89 & 0.14  \\
E15D & 0.22 & 0.74 & 0.13  \\
E15N & 0.53 & 1.07 & 0.57  \\
A16G & 1.06 & 1.09 & 0.82 \\
K17G & 0.38 & 2.32 & 0.80  \\
K18G & 0.7 &   0.99 & 0.75  \\
I20V & 0.4 &     1.3 &   0.14  \\
L21A & 0.25 & 1.33 &  -0.01  \\
L21G & 0.35 & 1.38 &  0.26  \\
D23A & -0.25 & 0.96 &  -0.41 \\
K24G & 0.1 &   3.19 &  0.12  \\
\hline
\end{tabular*}
\end{center}
\begin{flushleft}
Experimental $\Phi$-values and stability changes $\Delta G_N$ are from Itzhaki et al.\cite{Itzhaki95}. The change in intrinsic helix stability $\Delta G_\alpha$ is calculated with AGADIR \cite{Munoz95a,Munoz95b,Lacroix98}, see Merlo et al.~\cite{Merlo05}. The program AGADIR is based on helix/coil transition theory, with parameters fitted to data from Circular Dichroism (CD) spectroscopy. The free-energy changes are in units of kcal/mol. We only consider mutations with $\Delta G_N> 0.7$ kcal/mol, since $\Phi$-values for mutations with smaller values $\Delta G_N$ are often considered to be unreliable \cite{Fersht04,Garcia04,DelosRios06}.
\end{flushleft}
\end{table}

The central question is if we can reconstruct the transition state of a two-state protein from the observed $\Phi$-values for a large number of mutants \cite{Fersht99,Fersht04,Zarrine04,Raleigh05,Merlo05}. In the standard interpretation of $\Phi$-values, a $\Phi$-value of 1 is interpreted to indicate that the residue has a native-like structure in T, since the mutation shifts the free energy of the transition state T by the same amount as the free energy of the native state N. A $\Phi$-value of 0 is interpreted to indicate that the residue is as unstructured in T as in the denatured state D, since the mutation does not shift the free-energy difference between these two states. $\Phi$-values between 0 and 1 are typically taken to indicate partially native-like structure in T \cite{Fersht99,Fersht04}. In the traditional interpretation, a $\Phi$-value thus is taken to indicate the {\em degree of structure formation of the mutated residue in the transition-state ensemble T}.

However, the traditional interpretation is often not consistent. First, some $\Phi$-values are negative or larger than 1 \cite{Goldenberg99,Rios05} and cannot be interpreted as a degree of structure formation. An example is the negative $\Phi$-value $-0.25$ for the mutation D23A in the $\alpha$-helix of CI2 (see table 1). Second, $\Phi$-values are sometimes significantly different for different mutations at a given chain position. The mutations E15D and E15N in the helix of the protein CI2, for example, have $\Phi$-values of $0.22\pm 0.05$ and $0.53\pm 0.05$ \cite{Itzhaki95}, which differ by more than a factor 2 (see table 1). In the traditional interpretation, however, $\Phi$-values for different mutations of the same residue are expected to be identical, since they just reflect the degree of structure formation of this residue in T. Third, $\Phi$-values for neighboring residues within a given secondary structure often span a wide range of values. The $\Phi$-values shown in table 1 for mutations in the CI2 helix range from $-0.25$ to $1.06$. According to the traditional interpretation, this implies that some of the helical residues are unstructured in the transition state, while other residues, often direct neighbors, are highly structured. The traditional interpretation thus seems to contradict the notion that secondary structures are cooperative. In standard helix-coil models \cite{Zimm59,Chakrabartty95,Dill02}, the formation of helices requires that several consecutive helical turns are structured, stabilizing each other.

$\Phi$-values provide indirect information on the folding kinetics of a protein and, therefore, have attracted considerable theoretical interest. To understand the experimentally determined $\Phi$-values for a protein, molecular dynamics (MD) simulations with atomistic models are often performed 
\cite{Li94,Lazaridis97,Vendruscolo01,Li01,Paci02,Salvatella05,Daggett96,Day05,Settanni05,Lindorff03}. Such simulations are computationally demanding and in general do not allow direct calculations of folding rates and $\Phi$-values. Instead, the MD approaches typically rely on the assumption of the traditional interpretation that $\Phi$-values reflect the degree of structure formation of residues in the transition state T. For example, $\Phi$-values are often calculated from the fraction of contacts a residue forms in the transition state T, compared to the fraction of contacts in the native and the denatured states 
\cite{Li94,Lazaridis97,Vendruscolo01,Li01,Paci02,Guo03,Salvatella05}.  In an alternative approach, Daggett and coworkers compute an S-value \cite{Daggett96}, which is ``a measure of the amount of structure at a given residue, defined by the amounts of secondary and tertiary structure at each residue" \cite{Day05}. Exceptions to such structural assumptions are a recent MD study of an ultrafast mini-protein in which $\Phi$-values are calculated from rates for the wildtype and mutants via eq.~(\ref{in_phi_def}) \cite{Settanni05}, and the calculation of $\Phi$-values from free-energy shifts of the transition-state ensemble using eq.~(\ref{in_phi_trans_state_theo}) \cite{Lindorff03}. 

In the following sections, we will consider statistical-mechanical models that lead to a novel structural interpretation of mutational $\Phi$-values. The general conclusion from these models is that a consistent structural interpretation of $\Phi$-values (i) requires to split up mutation-induced stability changes into free-energy contributions from different substructural elements of a protein, and (ii) can be obtained with few parameters that characterize the degree of structure formation of cooperative substructures such as $\alpha$-helices and $\beta$-hairpins in the transition-state ensemble.

%%%%
\section{Formation of helices during protein folding}
%%%%
\label{section_model_helices}

In this section, we present a simple model for the formation of $\alpha$-helices during protein folding. The model will lead to a consistent structural interpretation of the mutational data for the CI2 helix shown in table 1 and for other helices. In particular, the model reproduces the negative $\Phi$-value for the mutation D23A in this helix, which cannot be understood in the traditional interpretation of $\Phi$-values (see last section). 

\begin{figure}
\begin{center}
\resizebox{0.6\linewidth}{!}{\includegraphics{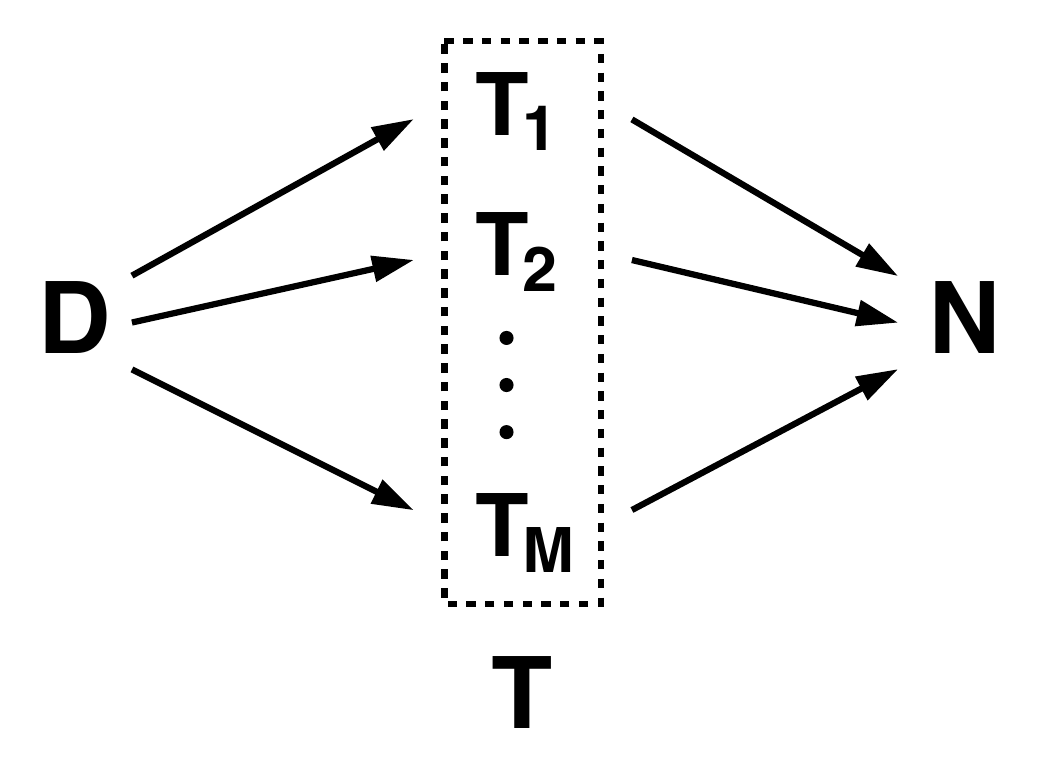}}
\end{center}
\caption{In our model, the transition-state ensemble T consists of $M$ transition-state conformations T$_1$, T$_2$, $\ldots$, T$_M$. The arrows indicate the folding direction from the denatured state D to the native state N via the transition-state conformations.}
\label{ts_model_helices}
\end{figure}

The model has two main ingredients. First, the central assumption is that a helix, or a segment of a helix, is either fully formed or not formed in partially folded conformations, in particular in transition-state conformations.  The transition state is described as an ensemble of $M$ different conformations (see fig.~\ref{ts_model_helices}). Each transition-state conformation is directly connected to the native state N and to the denatured state D. The model thus has $M$ parallel folding and unfolding routes. 

%Figure 

Second, mutation-induced free-energy changes are split into two components. The overall stability change $\Delta G_N$ is split into the change in intrinsic helix stability $\Delta G_\alpha$, and the free-energy change $\Delta G_t$ of tertiary interactions caused by the mutation: 
\begin{equation}
\Delta G_N=\Delta G_\alpha + \Delta G_t  \label{ts_helix_stability_decomposition}
\end{equation}
The intrinsic helix stability $G_\alpha$ is the stability of the `isolated' helix, i.e.~the free-energy difference between the folded and the unfolded state of the helix, in the absence of tertiary interactions with other structural elements. Similarly, we decompose each $\Delta G_m$, the mutation-induced free-energy change for the transition-state conformation $m$, into two terms:
\begin{equation}
\Delta G_m = s_m \Delta G_\alpha + t_m \Delta G_t \label{ts_components_ts}
\end{equation}
Here, $G_m$ is the free-energy difference between transition-state conformation $m$ and the denatured state.
Because we assume cooperative formation of the helix, or helical segment, $s_m$ is either 0 or 1, depending on whether the segment is formed or not in the transition-state conformation $m$. The coefficient $t_m$ is between 0 and 1 and represents the degree of tertiary structure formation in conformation $m$. 

We assume that the free-energy barrier for each transition-state conformation is significantly larger than the thermal energy, i.e.\ that $G_m/RT  \gg 1$ \cite{Schuler02,Akmal04}. The rate of folding along each route $m$ is then proportional to $\exp[-G_m/RT]$, and the total folding rate is the sum \cite{Weikl07}   
\begin{equation}
k \simeq c \sum_{m=1}^M e^{-G_m/RT}    \label{ts_folding_rate_helices}
\end{equation}
of the rates along the $M$ parallel routes. Here, $c$ is a constant prefactor.  

The folding rate for a mutant then is $k' =   k\big(G_1+\Delta G_1, G_2+\Delta G_2,\ldots, G_M+\Delta G_M\big)$ with $k$  given in eq.~(\ref{ts_folding_rate_helices}). We assume here that the mutations do not affect the pre\-factor $c$ in eq.~(\ref{ts_folding_rate_helices}). For small values $\left|\Delta G_m\right|$ of the mutation-induced free-energy changes, a Taylor expansion of $\ln k'$ leads to
\begin{eqnarray}
\ln k' -\ln k &\simeq&  \sum_{m=1}^M \frac{\partial \ln k_\text{wt}}{\partial G_m}\Delta G_m \nonumber \\
&=& -\frac{1}{RT}\frac{\sum_m  \Delta G_m e^{-G_m/RT}}{\sum_m e^{-G_m/RT}}
\end{eqnarray} 
With the decomposition of the $\Delta G_m$'s in eq.~(\ref{ts_components_ts}), we obtain
\begin{equation} 
\ln k' -\ln k \simeq  -\frac{1}{RT}\left(\chi_\alpha \Delta G_\alpha + \chi_t \Delta G_t \right) \label{ts_dlogk_helices}
\end{equation} 
with the two terms
\begin{equation}
\chi_\alpha  \equiv \frac{\sum_m s_m e^{-G_m/RT}}{\sum_m e^{-G_m/RT}}
\end{equation}
and
\begin{equation}
\chi_t  \equiv \frac{\sum_m t_m e^{-G_m/RT}}{\sum_m e^{-G_m/RT}} ~.
\end{equation}
The term $\chi_\alpha$ is the Boltzmann-weighted average of the secondary structure parameter $s_m$ in the transition-state ensemble T. The value $\chi_\alpha=1$ indicates that the helix is formed in all transition-state conformations $m$, while $\chi_\alpha=0$ indicates that the helix is formed in none of the transition-state conformations. Values of $\chi_\alpha$ between 0 and 1 indicate that the helix is formed in some of the transition-state conformation, and not formed in others. The term $\chi_t$ represents the Boltzmann-weighted average of the tertiary structure parameter $t_m$ in T. 

From eq.~(\ref{ts_dlogk_helices}) and the definition in eq.~(\ref{in_phi_def}), we then obtain the general form \cite{Weikl07}
\begin{equation}
\fbox{$\displaystyle
\Phi = \frac{\chi_\alpha\Delta G_\alpha + \chi_t\Delta G_t}{\Delta G_N} = \chi_t + \left(\chi_\alpha - \chi_t\right)\frac{\Delta G_\alpha}{\Delta G_N}$}
\label{phi_helices}
\end{equation}
of $\Phi$-values for mutations in helices. The second expression simply results from replacing $\Delta G_t$ by $\Delta G_N - \Delta G_\alpha$, see eq.~(\ref{ts_helix_stability_decomposition}). 

%Figure 
\begin{figure}[b]
\begin{center}
\resizebox{\linewidth}{!}{\includegraphics{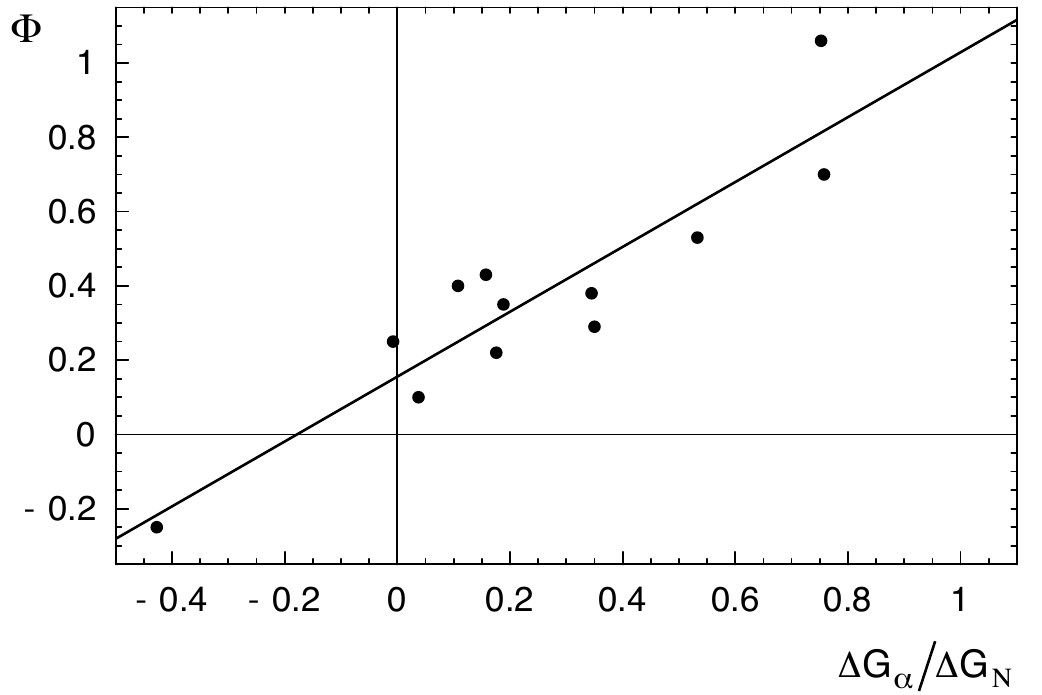}}
\end{center}
\vspace{-0.3cm}
\caption{Analysis of the mutational data for the $\alpha$-helix of CI2 shown in table 1. In agreement with eq.~(\ref{phi_helices}), we observe an approximately linear relation between $\Phi$ and $\Delta G_\alpha/\Delta G_N$ with a Pearson correlation coefficient of  0.91 \cite{Weikl07}. From the regression line $\Phi = 0.16 + 0.87 \Delta G_\alpha/\Delta G_N$, we obtain the structural parameters $\chi_\alpha = 1.03\pm 0.05$ and $\chi_t= 0.16\pm 0.05$. The structural parameter $\chi_\alpha$ close to 1 indicates that the helix is fully formed in the transition state, while the parameter $\chi_t$ indicates that tertiary interactions with the $\beta$-sheet are on average formed to a degree around 16 \%. The estimated standard deviation of data points from the regression line is 0.14 \cite{Weikl07}, which is comparable to the experimental errors \cite{Itzhaki95,Ruczinski06}.}
\label{figure_CI2helix}
\end{figure}
% 

%Figure 
\begin{figure}
\begin{center}
\resizebox{\linewidth}{!}{\includegraphics{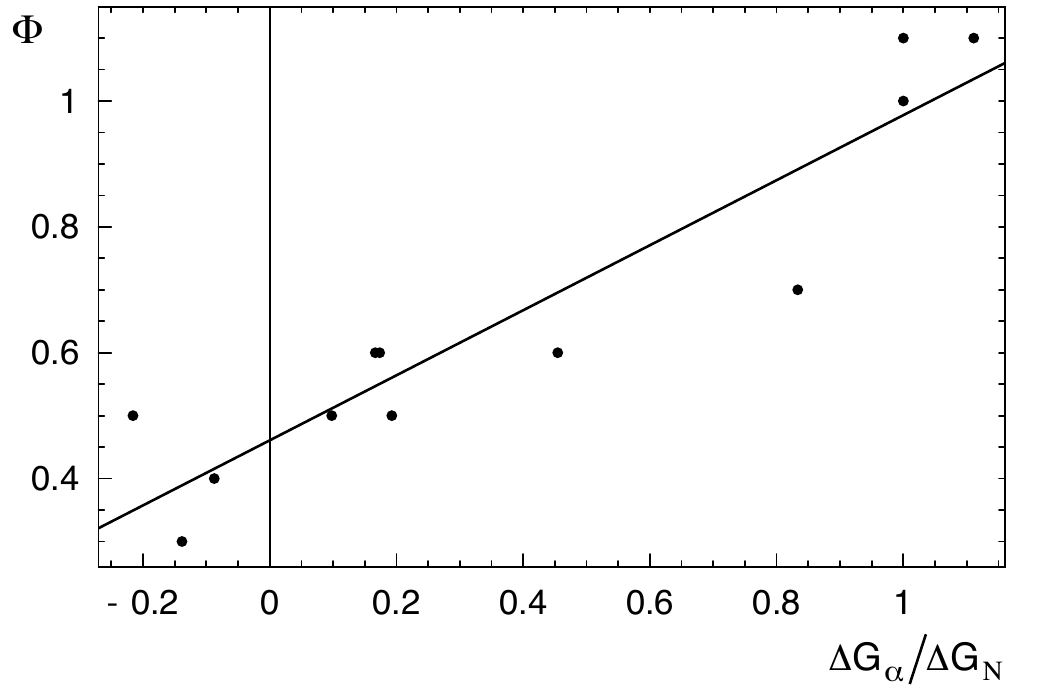}}
\end{center}
\vspace{-0.3cm}
\caption{Analysis of mutational data for helix 2 of protein A. The solid line represents the regression line $\Phi = 0.46 + 0.52 \,\Delta G_\alpha/\Delta G_N$. The Pearson correlation coefficient of the data points is 0.93, and the estimated standard deviation of the data points from the regression line is 0.10.  From the regression line and eq.~(\ref{phi_helices}), we obtain the structural parameters  $\chi_\alpha = 0.98 \pm 0.05$ and $\chi_t = 0.46\pm 0.05$. Values of $\Delta G_\alpha$ for the mutations have been estimated from a helix propensity scale \cite{Weikl07}.}
\label{figure_helix2_proteinA}
\end{figure}

The analysis of experimental $\Phi$-values and stability changes $\Delta G_N$ with eq.~(\ref{phi_helices}) requires an estimate of the mutation-induced changes $\Delta G_\alpha$ of the intrinsic helix stability. For the mutations in the CI2 helix shown in table 1, we have calculated $\Delta G_\alpha$ with the program AGADIR \cite{Merlo05}. In agreement with eq.~(\ref{phi_helices}), we observe a linear relation between $\Phi$ and $\Delta G_\alpha/\Delta G_N$ for the data shown in table 1, within reasonable errors (see fig.~\ref{figure_CI2helix}). The structural parameters $\chi_\alpha$ and $\chi_t$ can be estimated from the slope of the regression line, and the intersection of this line with the $y$-axis. For the CI2 helix, we obtain the values $\chi_\alpha = 1.03 \pm 0.05$ and $\chi_t = 0.16\pm 0.05$ \cite{Weikl07}, which implies that the helix is fully formed in the transition state, while tertiary interactions with the $\beta$-sheet are formed to an average degree of around 16 \%. 

In this model, the different $\Phi$-values for the mutations in the CI2 helix arise from different `free-energy signatures' $\Delta G_\alpha$ and $\Delta G_N$ of the mutations. In particular, the model captures the negative $\Phi$-value for the mutation D23A.  According to eq.~(\ref{phi_helices}), negative $\Phi$-values or $\Phi$-values larger than 1 can arise if  the mutation-induced changes $\Delta G_\alpha$ and $\Delta G_t= \Delta G_N- \Delta G_\alpha$ in secondary and tertiary free energy have opposite signs. We find that the mutation D23A stabilizes the helix ($\Delta G_\alpha<0$), but  destabilizes tertiary interactions ($\Delta G_t>0$).

The model leads to a consistent structural interpretation of the mutational data for several helices \cite{Weikl07}. Besides the CI2 helix, another helix for which a large number of mutational $\Phi$-values have been measured is helix 2 of the three-helix protein A. An analysis of the experimental data with eq.~(\ref{phi_helices}) leads to the structural parameters $\chi_\alpha = 0.98 \pm 0.05$ and $\chi_t = 0.46\pm 0.05$ (see fig.~\ref{figure_helix2_proteinA}). 
The value of $\chi_\alpha$ close to 1 indicates that the helix is fully formed in the transition state, and the value of $\chi_t$ close to 0.5 indicates that tertiary interactions with the other two helices of the protein are present to an average a degree of about 50 \%. 

%%%%
\section{Folding of small $\beta$-sheet proteins}
%%%%
\label{section_model_WW}

\begin{figure}[b]
\resizebox{\linewidth}{!}{\includegraphics{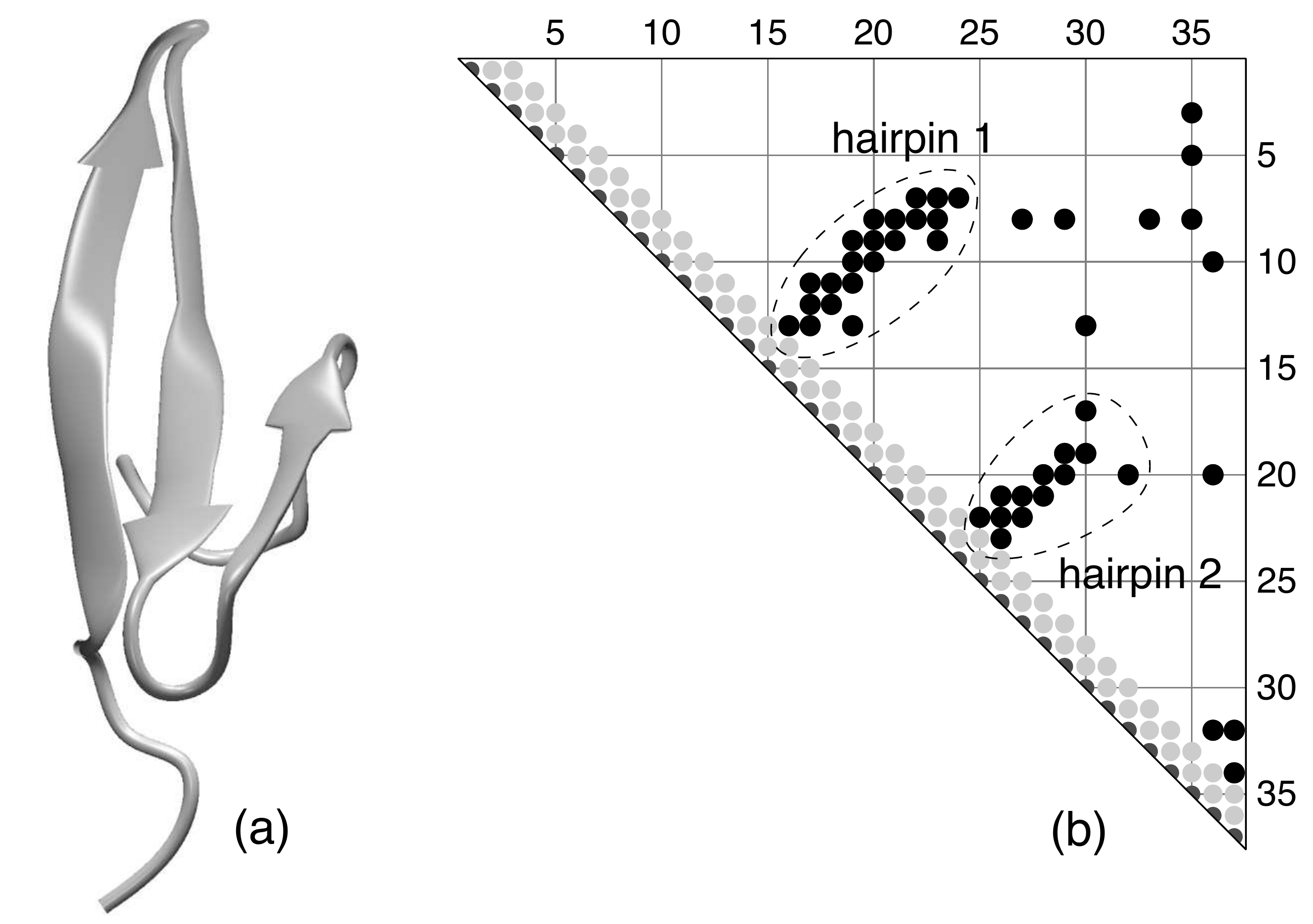}}
\caption{(a) The native structure of the FBP WW domain consist of two $\beta$-hairpins, which form a three-stranded $\beta$-sheet \cite{Macias00} . 
\label{figure_FBPWWstructure} 
(b) Contact matrix of the FBP domain. A black dot at position $(i,j)$ of the matrix indicates that the residues $i$ and $j$ are in contact. Two residues are defined here to be in contact if the distance between any of their non-hydrogen atoms is smaller than the cutoff distance 4 \AA. Contacts between nearest- and next-nearest neighboring residues are not considered (grey dots). The hairpins 1 and 2 of the WW domains correspond to clusters of contacts. The remaining contacts largely correspond to contacts of hydrophobic amino acids, the small hydrophobic core of the protein \cite{Weikl08b}. 
}
\end{figure}

In this section, we model mutational data for the folding dynamics of small $\beta$-sheet proteins. The smallest $\beta$-proteins have just three $\beta$-strands. Important representatives of this class of proteins are WW domains (see fig.~\ref{figure_FBPWWstructure}), named after two conserved tryptophan residues, which are represented by the letter W in the single-letter code for amino acids. WW domains are central model systems for understanding $\beta$-sheet folding and stability \cite{Jaeger01,Ferguson02,Deechongkit04,Socolich05,Petrovich06}.

The fastest three-stranded $\beta$-proteins fold in microseconds and are, thus, good targets for MD folding simulations with atomistic models (see section \ref{section_folding_dynamics}). For a small, designed three-stranded $\beta$-sheet protein, beta3s, the transition-state conformations have been determined from extensive folding-unfolding MD simulations \cite{Rao05}. The native structure of beta3s is similar to the structure of WW domains, with two $\beta$-haipins forming an antiparallel three-stranded $\beta$-sheet. By identifying clusters of structurally similar conformations that have the same probability to fold or unfold, Rao et al.\ \cite{Rao05} obtained a transition-state ensemble for beta3s in which either hairpin 1 or hairpin 2  is structured, while the other hairpin is unstructured.  The two $\beta$-hairpins of beta3s thus appear to be cooperative substructures that are fully structured or unstructured in the transition state. 

%Figure 
%

%Figure 
\begin{figure}[t]
\begin{center}
\resizebox{0.65\linewidth}{!}{\includegraphics{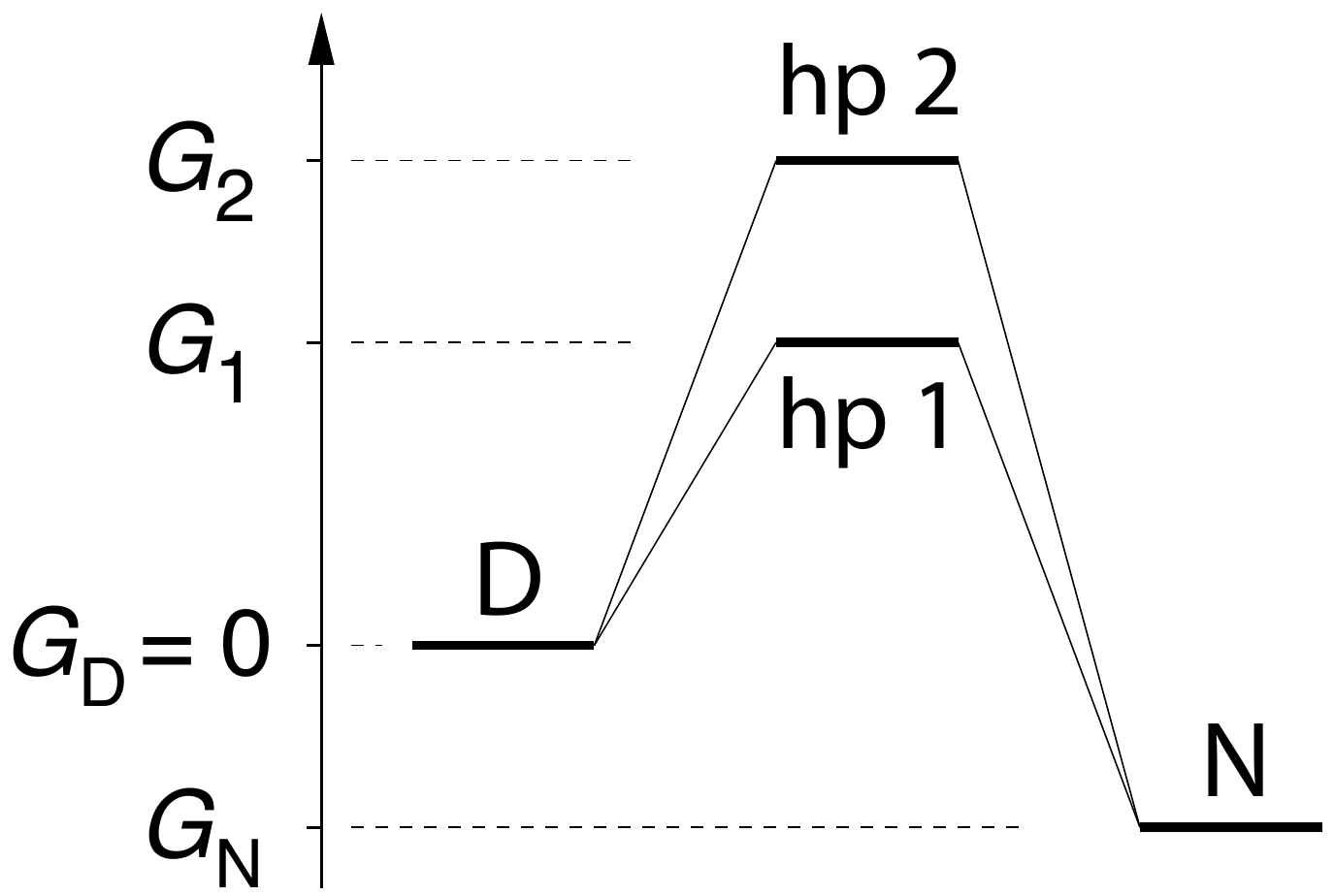}}
\end{center}
\caption{Simple energy landscape of the four-state model for three-stranded $\beta$-sheet proteins. The four states are the denatured state D, the native state N, and two partially folded states hp 1 and hp 2 in which one of the two hairpins is formed. Here, $G_N$ is the free-energy difference between the native state N and the denatured state D, which has the `reference free energy' $G_D = 0$, and  $G_1$ and $G_2$ are the free-energy differences between the transition-state conformations and the denatured state.}
\label{figure_WWlandscape} 
\end{figure}

In the statistical-mechanical model for three-stranded $\beta$-sheet proteins considered here, we assume a beta3s-like transition-state ensemble for in which either hairpin 1 or hairpin 2 are formed (see fig.~\ref{figure_WWlandscape}). The model has two folding routes: On one of the routes, hairpin 1 forms before hairpin 2, and on the other route, after hairpin 2. The energy landscape of this model can be characterized by three free-energy differences: The free-energy difference $G_N$ of the native state and the free-energy differences $G_1$ and $G_2$ of the two transition-state conformations with respect to the denatured state (see fig.~\ref{figure_WWlandscape}). For large transition-state barriers $G_1$ and $G_2$, the folding rate is \cite{Weikl08b}
\begin{equation}
k \simeq  c \left(e^{-G_1/RT} +e^{-G_2/RT}\right) 
\label{ts_folding_rate_WW}
\end{equation}
The folding rate $k$ is the sum of the rates for the two folding routes. 

Mutations correspond to perturbations of the free-energy landscape. In this model, a mutation can be characterized by the free-energy changes $\Delta G_1$, $\Delta G_2$, and $\Delta G_N$. The folding rate of the mutant then is $k' = k(G_1+\Delta G_1, G_2+\Delta G_2)$. For small perturbations $\Delta G_1$ and $\Delta G_2$, a Taylor expansion of $\ln k'$ to first order leads to
\begin{eqnarray}
\ln k' - \ln k &\simeq&  \frac{\partial \ln k}{\partial G_1} \Delta G_1 +  \frac{\partial \ln k}{\partial G_2} \Delta G_2  \nonumber \\
&=& -\frac{1}{RT} \left(\chi_1 \Delta G_1 + \chi_2\Delta G_2\right) \label{ts_dlogk_WW}
\end{eqnarray}
with
\begin{equation}
\chi_1 \equiv \frac{e^{-G_1/R T}}{e^{-G_1/R T} + e^{-G_2/RT}} 
\end{equation}
and
\begin{equation}
\chi_2 \equiv \frac{e^{-G_2/R T}}{e^{-G_1/R T} + e^{-G_2/RT}}
\end{equation}
The two parameters $\chi_1$ and $\chi_2$ are the probabilities that conformation 1 with hairpin 1 and conformation 2 with hairpin 2 are populated in the transition-state ensemble. From the $\Phi$-value definition (\ref{in_phi_def}) and eq.~(\ref{ts_dlogk_WW}), we obtain the general form \cite{Weikl08b}
\begin{equation}
\fbox{$\displaystyle
\Phi = \frac{\chi_1 \Delta G_1 + \chi_2 \Delta G_2}{\Delta G_N}$}
 \label{phi_WW}
\end{equation}
of $\Phi$-values for mutations in three-stranded $\beta$-sheet proteins. 

%Figure 
\begin{figure}[b]
\resizebox{\linewidth}{!}{\includegraphics{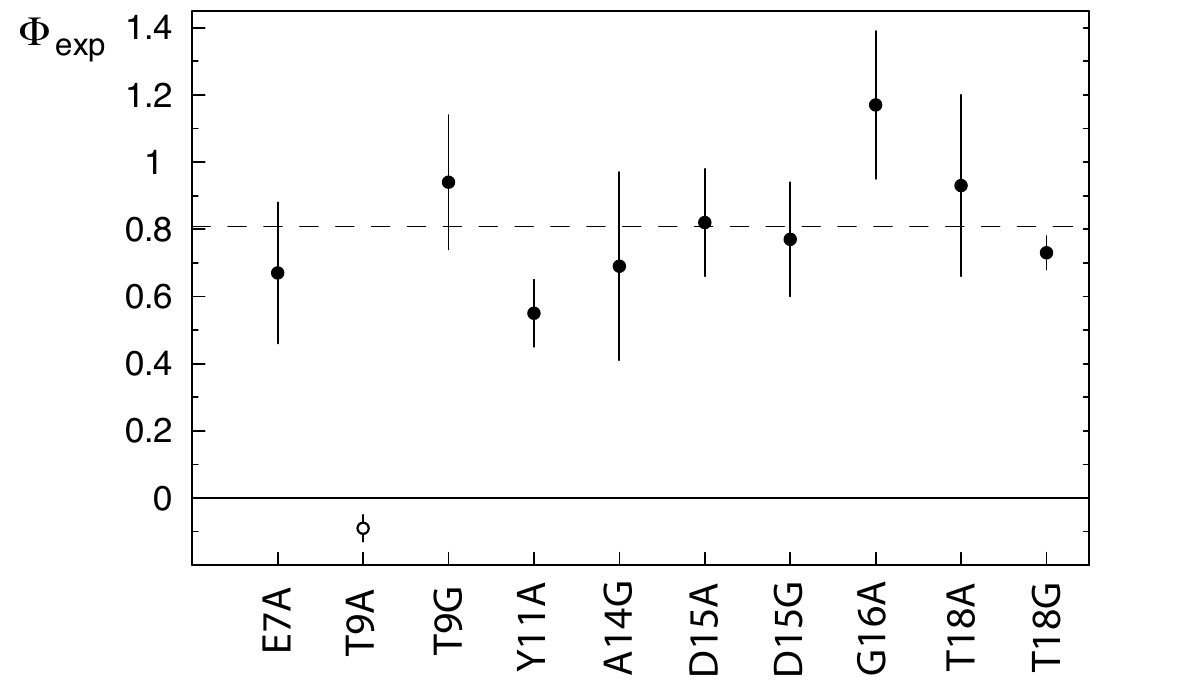}}
\caption{$\Phi$-values for mutations that only affect haipin 1 of the FBP WW domain \cite{Petrovich06}. Except for one outlier (open circle for mutation T9A), the $\Phi$-values are centered around the mean value $0.81\pm 0.06$, with deviations mostly within the experimental errors. 
}
\label{FBPphis_hp1}
\end{figure}

A detailed mutational analysis of the folding kinetics of the FBP WW domain shown in fig.~\ref{figure_FBPWWstructure} has been performed by Petrovich et al.~\cite{Petrovich06}. In general, mutations can affect hairpin 1, hairpin 2, or the small hydrophobic core of the protein. Interestingly, eq.~(\ref{phi_WW}) predicts that all mutations that affect, e.g., only hairpin 1 should have the same $\Phi$-value $\chi_1$ since we have $\Delta G_2 = 0$ and $\Delta G_N = \Delta G_1$ for these mutations. This is indeed the case, except for one outlier (see fig.~\ref{FBPphis_hp1}). The $\Phi$-values of the remaining nine mutations that affect only hairpin 1 of the FBP domain are centered around the mean value 0.81 (dashed line in fig.~\ref{FBPphis_hp1}), mostly within experimental errors. The mean value of these nine $\Phi$-values leads to the estimate $\chi_1=0.81\pm 0.06$ \cite{Weikl08b}. Similarly, the four $\Phi$-values for mutations that affect only hairpin 2 are centered around a mean value $\chi_2=0.30 \pm 0.08$ \cite{Weikl08b}. Within the statistical errors, these two estimates for $\chi_1$ and $\chi_2$ sum up to 1, which is a consistency requirement of our model since the protein has to take one of the two possible routes to the native state (see fig.~\ref{figure_WWlandscape}). The two parameters $\chi_1$ and $\chi_2$ are the probabilities for the two routes. 

\begin{figure}[t]
\resizebox{\linewidth}{!}{\includegraphics{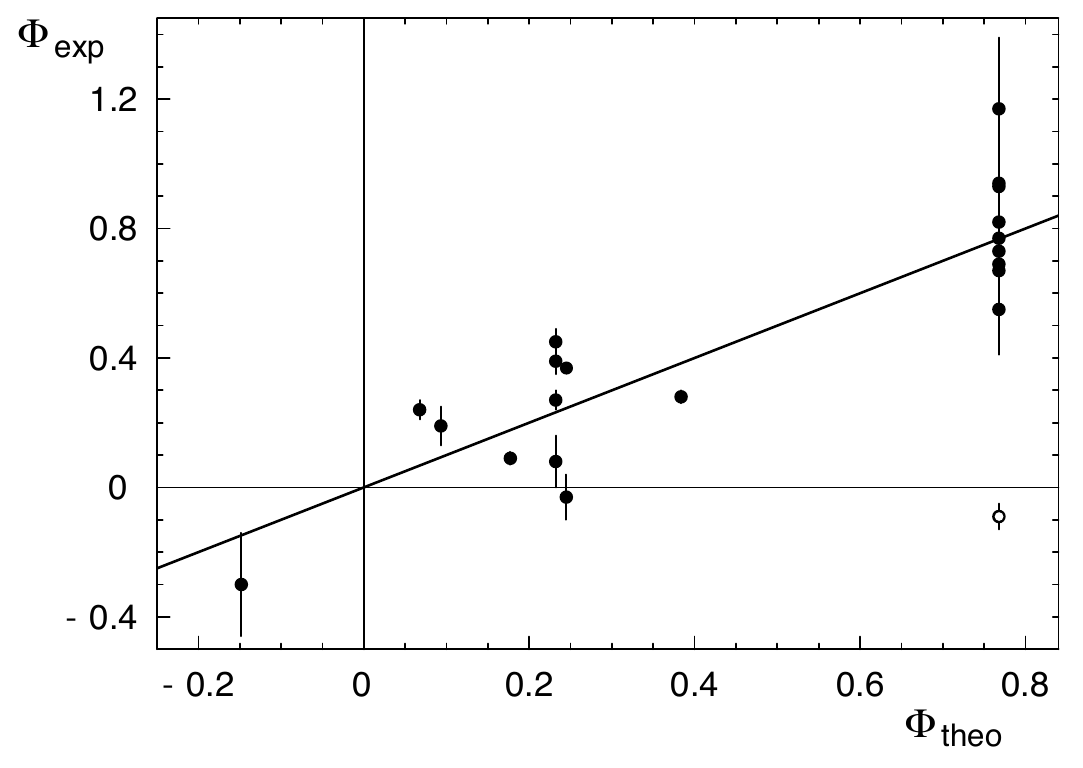}}
\caption{Experimental versus theoretical $\Phi$-values for the FBP WW domain. The theoretical $\Phi$-values have been obtained from a least-square fit of eq.~(\ref{phi_WW}) with the single fit parameter $\chi_1$. From this fit, we obtain the values $\chi_1=0.77\pm 0.05$ and $\chi_2=1-\chi_1=0.23\pm 0.05$ for the fractions of the two transition-state conformations in which either hairpin 1 or hairpin 2 are formed. The Pearson correlation coefficient between theoretical and experimental $\Phi$-values is 0.90 if the outlier data point for mutation T9A (open circle) is not considered, and 0.77 if the outlier is included \cite{Weikl08b}.
}
\label{figure_FBPphis}
\end{figure}

To include other mutations in the model, we have to estimate the impact of these mutations on the stability of the different structural elements (hairpin 1, hairpin2, or the hydrophobic core) they affect. We have used the program FOLD-X \cite{Guerois02,Schymkowitz05} to calculate these stability changes \cite{Weikl08b}. The structural parameters $\chi_1$ and $\chi_2$ then can be obtained from a least-square fit of eq.~(\ref{phi_WW}) to the experimental data (see fig.~\ref{figure_FBPphis}), with a single fit parameter since $\chi_1+\chi_2=1$.  The structural information obtained from this fit  is that the transition-state ensemble of the FBP WW domain consists to roughly $\frac{3}{4}$ of conformation 1 with hairpin 1 formed, and to $\frac{1}{4}$ of conformation 2 with hairpin 2 formed. 

%Figure 

In this model, the magnitude of a $\Phi$-value depends on which structural elements are affected, and on the mutation-induced free-energy changes of these elements. As in the previous section, negative $\Phi$-values or $\Phi$-values larger than 1 can arise if a mutation has both stabilizing and destabilizing effects on different structural elements. For example, the model reproduces the negative $\Phi$-value $-0.30$ for a mutation of the FBP WW domain that stabilizes hairpin 2 but destabilizes the hydrophobic core (see fig.~\ref{figure_FBPphis}), according to calculations with the program FOLD-X. The model also leads to a consistent interpretation of $\Phi$-values for the PIN WW domain \cite{Jaeger01,Deechongkit04} with the structural parameters $\chi_1=0.67\pm 0.05$ and $\chi_2= 0.33\pm 0.05$ \cite{Weikl08b}.

The deviations between experimental and theoretical $\Phi$-values in fig.~\ref{figure_FBPphis} are mostly within reasonable errors. It has been recently suggested that experimental errors for $\Phi$-values  may be underestimated since it is usually assumed that the errors in the measured free-energy changes of the transition state and the folded state are independent, which is not the case \cite{Ruczinski06} (see also refs.~\cite{DelosRios06,Sanchez03,Fersht04,Garcia04,Cobos08} for a discussion on experimental errors of $\Phi$-value measurements). 
Other sources of errors are the simplifying modeling assumptions on the transition-state structure, and the calculations of the mutation-induced free-energy changes.

In a related approach, Zarrine-Afsar et al.~\cite{Zarrine07} have found that the folding rate changes for different mutations of the same residue in the $\beta$-sheet of the Fyn SH3 domain correlate with changes in $\beta$-sheet propensity, a simple measure for mutation-induced free-energy changes in the $\beta$-sheet. More recently, Farber and Mittermaier \cite{Farber08} have modeled the effects of different mutations of hydrophobic core residues with two structural parameters for hydrophobic burial and native-like interactions.

%%%%
\section{Discussion and Conclusions}
%%%%

We have considered the question how transition states of two-state protein folding can be reconstructed from mutational data for the folding dynamics. In the traditional interpretation of the mutational data, the structural parameters are the degrees of structure formation of each residue of the protein in the transition state. The number of structural parameters thus is identical with the number of residues. In this interpretation, the $\Phi$-values for mutations of a given residue are taken to be identical with the residue's degree of structure formation in the transition state  (see section \ref{section_mutational_analysis}), which can lead to inconsistencies:  The traditional interpretation cannot capture different $\Phi$-values for different mutations of the same residue, and `non-classical' $\Phi$-values smaller than 0 or larger than 1. 

In sections \ref{section_model_helices} and \ref{section_model_WW}, we have considered a different structural interpretation of $\Phi$-values for mutations in $\alpha$-helices and small $\beta$-sheet proteins. This novel interpretation implies just two structural parameters per helix, the degrees of secondary and tertiary structure of the helix in the transition state, and a single structural fitting parameter for three-stranded  $\beta$-proteins, the relative degree of structure formation of hairpin 1 and hairpin 2 in the transition state. Inconsistencies of the traditional interpretation are resolved by splitting mutation-induced free-energy changes into secondary and tertiary components. In particular, two negative $\Phi$-values for a mutation in the CI2 helix and a mutation in the FBP WW domain are traced back to free-energy changes of opposite sign, without additional assumptions. The mutations stabilize the CI2 helix and hairpin 2 of the FBP WW domain, respectively, but destabilize tertiary interactions with other structural elements of the proteins. Other groups have suggested that negative $\Phi$-values may arise from non-native interactions in the transition state \cite{Li00}, parallel folding routes with energetic traps \cite{Ozkan01}, experimental errors \cite{Sanchez03}, or from mutation-induced free-energy changes of the denatured state \cite{Cho06}. 
An extension of the novel interpretation to larger $\beta$-sheet proteins than the three-stranded WW domains considered here requires the identification of cooperative substructural elements. 
Candidates for such cooperative elements are $\beta$-hairpins or other $\beta$-strand pairings \cite{Reich06}. 

Future MD folding simulations with detailed atomistic models may lead to a more complete understanding of protein folding transition states and mutational effects on the folding dynamics. Challenging goals are the characterization of transition-state conformations on folding or unfolding trajectories  \cite{Rao05} and the direct determination of $\Phi$-values from folding simulations with mutants  \cite{Settanni05}.

%%%% Acknowledgments %%%%%%%%
\section*{Acknowledgments}

The author would like to thank Ken Dill for numerous discussions and joint work on this subject.

\end{document}